\documentclass[fleqn]{article}
\usepackage{espcrc2}
\usepackage{graphicx}
\usepackage[figuresright]{rotating}
\usepackage{psfig}

\newcommand{\chicJ}{\chi_{cJ}}

\newcommand{\ppp}{\pi^+\pi^- \pi^0}

\newcommand{\GG}{\gamma\gamma}

\newcommand{\jpsi}{J/\psi}
\newcommand{\ar}{\rightarrow}

\newcommand{\ww}{\omega\omega}

\newcommand{\bfg}{\begin{figure}}
\newcommand{\efg}{\end{figure}}
\newcommand{\bitm}{\begin{itemize}}
\newcommand{\eitm}{\end{itemize}}
\newcommand{\bnum}{\begin{enumerate}}
\newcommand{\enum}{\end{enumerate}}
\newcommand{\btbl}{\begin{table}}
\newcommand{\etbl}{\end{table}}
\newcommand{\btbu}{\begin{tabular}}
\newcommand{\etbu}{\end{tabular}}
\newcommand{\bcl}{\begin{center}}
\newcommand{\ecl}{\end{center}}
\newcommand{\bbt}{\bibitem}
\newcommand{\beq}{\begin{equation}}
\newcommand{\eeq}{\end{equation}}
\newcommand{\beqr}{\begin{eqnarray}}
\newcommand{\eeqr}{\end{eqnarray}}

\begin{document}

\title{Observation of $\chicJ \to \ww$ decays}
\author{\small{M.~Ablikim$^{1}$,        J.~Z.~Bai$^{1}$,  Y.~Ban$^{11}$,
J.~G.~Bian$^{1}$,        X.~Cai$^{1}$,
H.~F.~Chen$^{16}$,
H.~S.~Chen$^{1}$,        H.~X.~Chen$^{1}$,
J.~C.~Chen$^{1}$,
Jin~Chen$^{1}$
% ...
\vspace{0.2cm}\\
(BES Collaboration)\\
\vspace{0.2cm}
{\it
$^{1}$ Institute of High Energy Physics, Beijing 100049, People's
Republic of China\\
% ...
$^{11}$ Peking University, Beijing 100871, People's Republic of China\\
% ...
$^{16}$ University of Science and Technology of China, Hefei 230026,
People's Republic of China\\
% ...
}}}
\date{\today}  
\begin{abstract}
Decays of $\chi_{c0,2}\ar\ww$ are observed for the first time using a
sample of $14.0\times 10^6$ $\psi(2S)$ events collected with the BESII
detector. The branching ratios are determined to be ${\cal
B}(\chi_{c0}\ar \ww)=(2.29\pm 0.58\pm 0.41)\times 10^{-3}$ and ${\cal
B}(\chi_{c2}\ar\ww)=(1.77\pm 0.47\pm 0.36)\times 10^{-3}$, where the
first errors are statistical and the second systematic. The
significances of the two signals are $4.4\sigma$ and $4.7\sigma$,
respectively.
\end{abstract}

\maketitle

\section{Introduction}
Exclusive quarkoninum decays provide an important laboratory for
investigating perturbative quantum chromodynamics. Compared with
$\jpsi$ and $\psi(2S)$ decays, one has much less knowledge on
$PC=++\chicJ$ decays. While a few exclusive decays of $\chicJ$ have
been measured, many decay modes remain unknown. Current theoretical
analyses of $\chicJ$ decays provide only a rough treatment of the
color-octet wave function. For $\chicJ\ar vector~vector$ mode, so far
only measurements of $\chicJ\ar\phi\phi$~\cite{2phi} and
$\chicJ\ar K^*(892)^0\bar{K}^*(892)^0$ ~\cite{kstar} are available
with low statistics.  Precise measurements for more channels
will help in better understanding the various mechanism~\cite{color,hrb}
of $\chicJ$ decays and the nature of $^3P_J~c\bar{c}$ bound states.

  Further, the decays of $\chicJ$, especially $\chi_{c0}$ and
$\chi_{c2}$, provide a direct window on glueball dynamics in the
$0^{++}$ and $2^{++}$ channels since the hadronic decays may proceed
via $c\bar{c}\ar gg\ar q\bar{q}q\bar{q}$.

Recently, the branching ratio for $\chi_{c0}\ar
f_0(980)f_0(980)$~\cite{f0} has been measured by the BES
collaboration.  In the present analysis, a search for $\chi_{c0,2}$
decaying into $\ppp\ppp$ final states is carried out using 14 million
$\psi(2S)$ events~\cite{moxh} accumulated at the upgraded BES detector
(BESII). Signals of $\chi_{c0}$ and $\chi_{c2}$ decaying to $\omega$
pairs in $\psi(2S)$ radiative decays are observed for the first time.
\section{The BES detector}
The Beijing Spectrometer (BES) is a conventional solenoidal magnet
detector that is described in detail in Ref.~\cite{bes}; BESII is the
upgraded version of the BES detector~\cite{bes2}. A 12-layer vertex
chamber (VC) surrounding the beam pipe provides trigger
and position information. A forty-layer main drift chamber (MDC),
located radially outside the VC, provides trajectory and energy loss
($dE/dx$) information for charged tracks over $85\%$ of the total
solid angle.  The momentum resolution is $\sigma _p/p = 0.017
\sqrt{1+p^2}$ ($p$ in $\hbox{\rm GeV/c}$), and the $dE/dx$ resolution
for hadron tracks is $\sim 8\%$.  An array of 48 scintillation
counters surrounding the MDC measures the time-of-flight (TOF) of
charged tracks with a resolution of $\sim 200$ ps for hadrons.
Outside of the TOF counters is a 12-radiation-length barrel shower
counter (BSC) comprised of gas proportional tubes interleaved with
lead sheets. This measures the energies of electrons and photons over
$\sim 80\%$ of the total solid angle with an energy resolution of
$\sigma_E/E=22\%/\sqrt{E}$ ($E$ in GeV).  Outside of the solenoidal
coil, which provides a 0.4~Tesla magnetic field over the tracking
volume, is an iron flux return that is instrumented with three double
layers of counters that identify muons of momentum greater than 0.5
GeV/c.

A GEANT3 based Monte Carlo (MC) program with
detailed consideration of the detector performance (such as dead
electronic channels) is used to simulate the BESII detector.  The
consistency between data and Monte Carlo has been carefully checked in
many high purity physics channels, and the agreement is quite
reasonable~\cite{simbes}.

\section{Event selection}
\subsection{\boldmath $\ww$ signal}
In this analysis, $\chicJ \to \omega \omega \to \ppp \ppp$ channels are
investigated using $\psi(2S)$ radiative decays to  $\chicJ$.
Events with four charged tracks
and five or six photons are selected. 
Each charged track is required to be well fit by a three-dimensional
helix and to have a polar angle, $\theta$, within the fiducial region
$|\cos\theta|<0.8$. To ensure tracks originate from the interaction
region, we require $V_{xy}=\sqrt{V_x^2+V_y^2}<2$ cm and $|V_z|<20$ cm,
where $V_x$, $V_y$, and $V_z$ are the $x, y$ and $z$ coordinates of the
point of closest approach of each track to the beam axis.

A neutral cluster is considered to be a photon candidate if it is
located within the BSC fiducial region ($|\cos\theta|<0.8$), the
energy deposited in the BSC is greater than 40 MeV, the first hit
appears in the first 10 radiation lengths, and the angle between the
cluster and the nearest charged track is greater than $6^\circ$.

A six constraint (6-C) kinematic fit to the hypothesis
$\psi(2S)\ar\gamma\ppp\ppp$ with the invariant mass of the two photon pairs
constrained to the $\pi^0$ mass is performed, and the $\chi^2$
of the 6-C fit is required to be less than 15. For events with six
photons candidates, the combination having the minimum $\chi^2$ is
chosen, and the probability of the 6-C fit is required to be larger than
that of the 7-C fit to the hypothesis $\psi(2S)\ar 2\pi^+2\pi^-
3\pi^0$ to suppress potential background from 
$\psi(2S)\ar\omega\pi^+\pi^-\pi^0\pi^0\ar 2\pi^+2\pi^-3\pi^0$.

Since there are four $\omega$ pair combinations from $\ppp\ppp$,
the $\omega$ pair with the minimum $R$, which is defined as
$$R=\sqrt{(M_{\ppp}^1-0.783)^2+(M_{\ppp}^2-0.783)^2},$$
is chosen for further analysis. Here, $M_{\ppp}$ is the invariant mass
of three pions and superscript 1, 2 denote different pion
combinations. Therefore, there is only one entry for each 
event.

\begin{figure}[!ht]
\centerline{\psfig{file=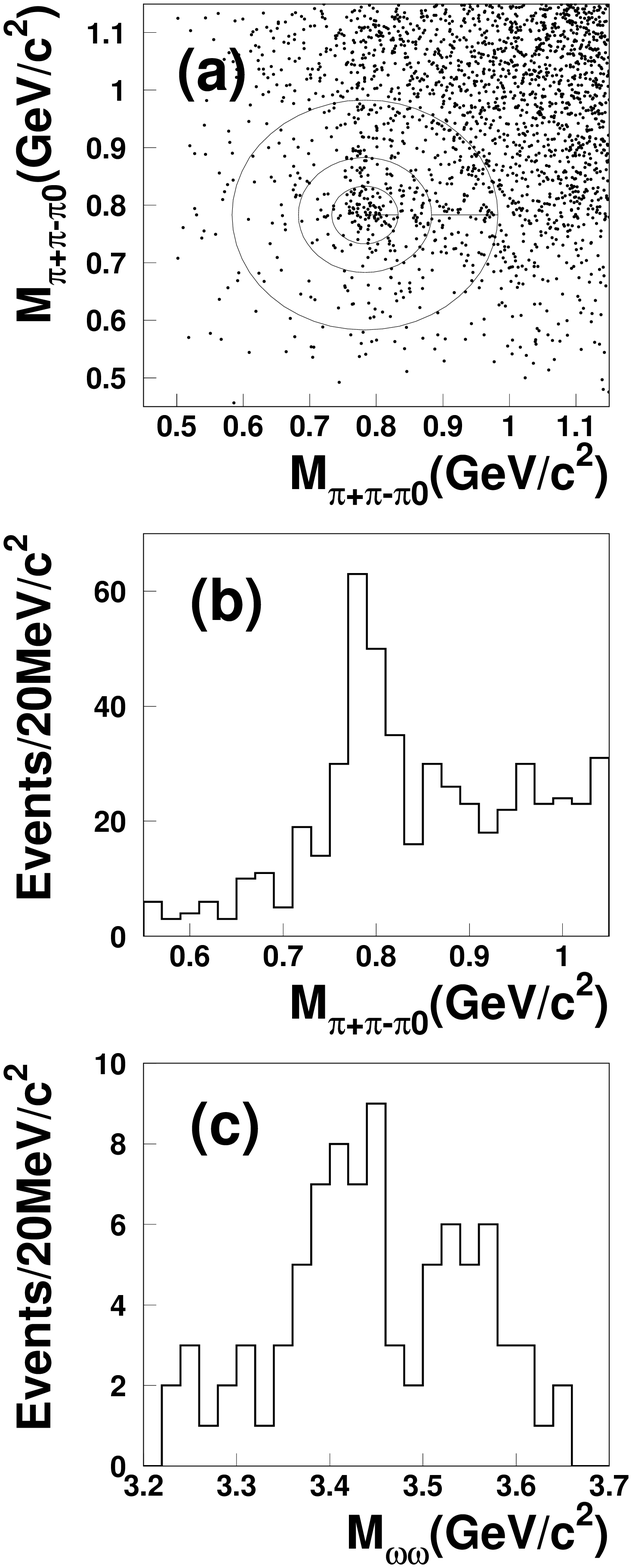,width=5cm,height=12cm}}
\caption{Distributions of events surviving the selection criteria
described in the text with $M_{6\pi}>3.2$ GeV$/c^2$. (a) $M_{\ppp}$
versus $M_{\ppp}$, (b) $M_{\ppp}$ recoiling against the opposite
$\omega$, selected by requiring $|M_{\ppp}-783|<50$ MeV/$c^2$, and (c)
$M_{\ww}$ invariant mass distribution for events where the $\omega$
pair satisfies $R<50$ MeV/$c^2$.}
\label{gww}
\end{figure}

\begin{figure}[!ht]
\hspace{-0.5cm}
\centerline{\psfig{file=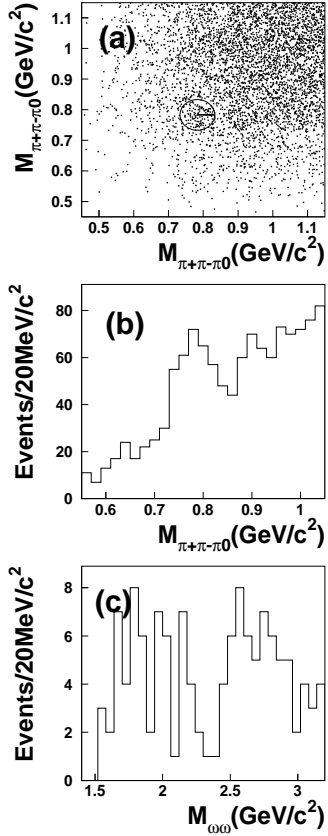,width=5cm,height=12cm}}
\caption{Distributions defined as in Fig.~\ref{gww} but with $M_{6\pi}<3.2$ GeV/$c^2$.}
\label{gww32}
\end{figure}

Figures~\ref{gww} and ~\ref{gww32} show mass distributions for
candidate events in the high mass  
($M_{6\pi}>3.2$ GeV/$c^2$) and low mass regions ($M_{6\pi}<3.2$ GeV/$c^2$), 
respectively. 
Here (a) is the scatter plot of $M_{\ppp}$ versus $M_{\ppp}$, (b) 
is the  
$M_{\ppp}$ distribution recoiling against the opposite $\omega$,
selected by requiring  
$|M_{\ppp}-783|<50$ MeV/$c^2$, and (c) is the  $M_{\ww}$ invariant
mass distribution for events in the $\omega$ pair signal region,
defined by $R<50$ MeV/$c^2$.
In Fig.~\ref{gww}, clear $\omega$ 
signal can be seen in (b), and clear $\chi_{c0}$ and $\chi_{c2}$
signals in (c),
indicating the existence of $\chi_{c0,2}\ar\ww$ decays. By contrast,
in the low $M_{6\pi}$ mass 
region, shown in Fig.~\ref{gww32}, the $\omega$ pair signal 
is less significant than in the high mass region. Here, only $\omega$ 
pair events in high mass region are studied. 

In order to test if the selection criteria in this analysis will give
`fake' $\omega$ pair events from non-$\omega$ pair events, 300000 MC
simulated $\psi\ar\gamma\chi_{c0}\ar\gamma 6\pi$ events are generated
in which $\chi_{c0}\ar 6\pi$ decays according to the phase space.
Fig.~\ref{ps} shows the $M_{\ppp}$ distributions of the surviving MC
phase space events after requiring the same selection criteria as for the
real data. No peak around the $\omega$ mass is seen, which shows that the 
$\omega$ pair selection
criteria in this analysis does not generate fake $\omega$ pair signals.
\bfg[ht]
\centerline{\psfig{file=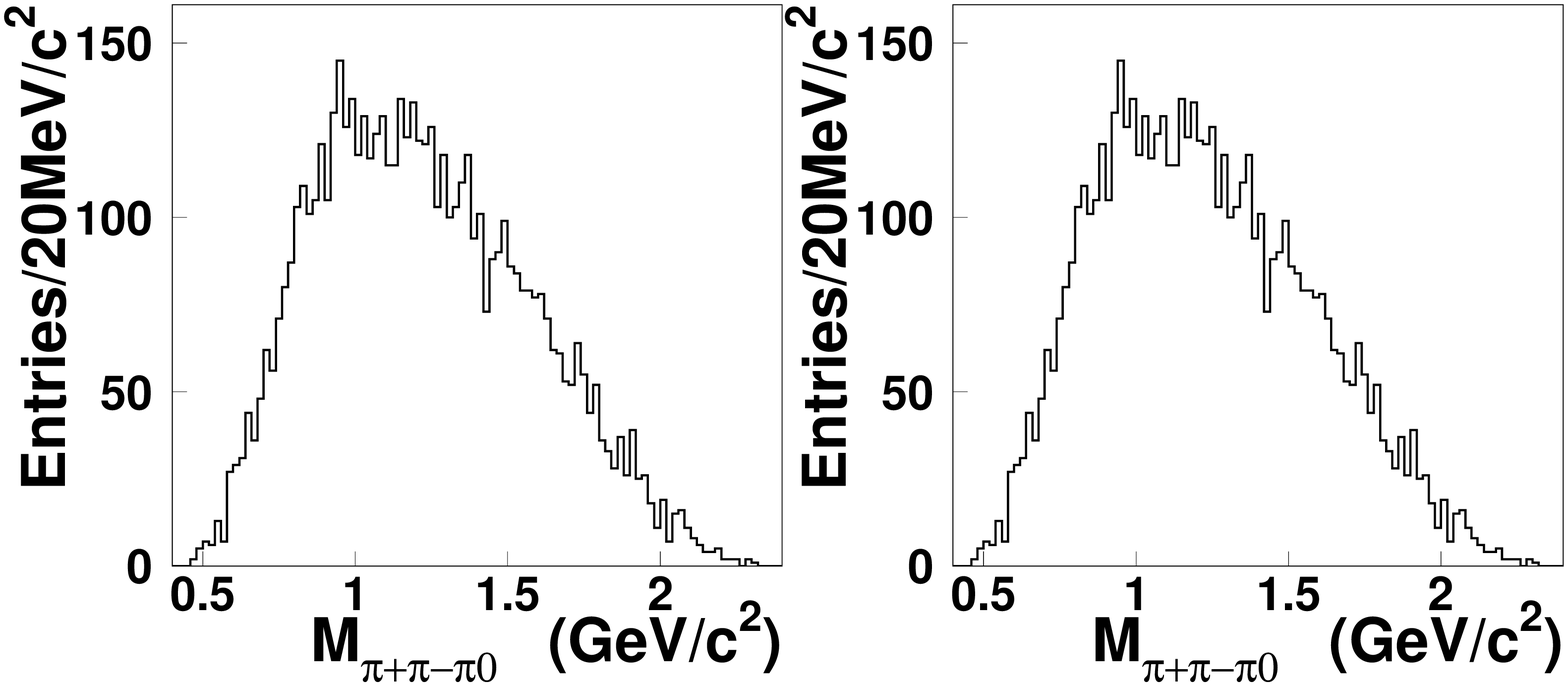,width=8cm,height=4cm}}
\caption{$M_{\ppp}$ distributions from MC phase space simulated  
$\psi(2S)\ar\gamma\chi_{c0},~\chi_{c0}\ar\ppp\ppp$.}
\label{ps}
\end{figure}

\begin{figure}[htpb]
\centerline{\psfig{file=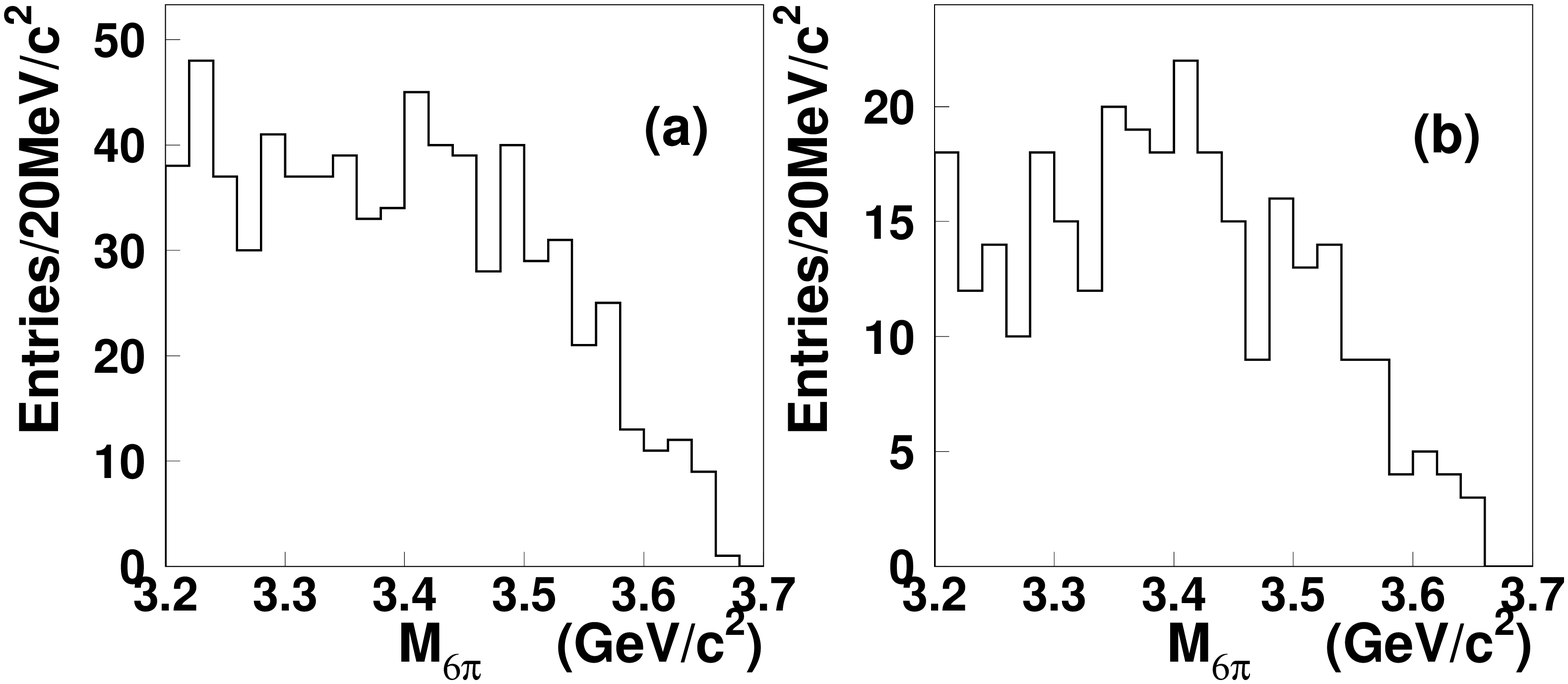,width=8cm,height=4cm}}
\caption{$M_{6\pi}$ distribution of events in sideband regions 
(a)~$150<R<300$ ~MeV$/c^2$ and (b)~$100<R<200$ ~MeV$/c^2$.}
\label{sb}
\end{figure}

The annular region around the $\omega$ pair signal circle, shown in
Fig~\ref{gww}(a), is taken as the sideband region. Fig.~\ref{sb} shows
the $M_{6\pi}$ sideband distributions defined using the radius R to be (a)
$150<R<300$ MeV$/c^2$  and (b) $100<R<200$ MeV/$c^2$. No obvious $\chicJ$ 
signals seen in these sideband distributions.

\subsection{MC simulation}
\begin{figure}[ht]
\vspace{-0.5cm}
\hspace{-.5cm}
\centerline{\psfig{file=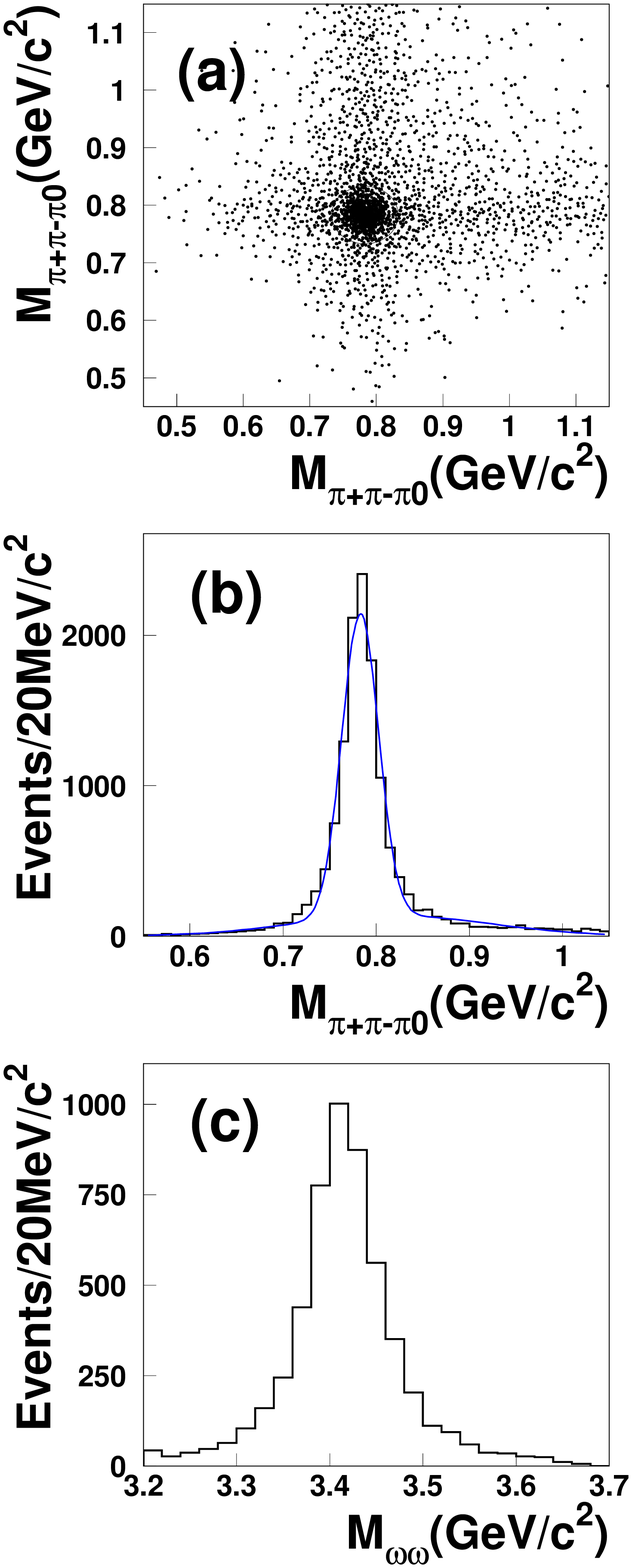,width=5cm,height=12cm}}
\caption{Distributions defined as in Fig.~\ref{gww} from
MC simulated $\psi(2S)\ar\gamma\chi_{c0},~\chi_{c0}\ar\omega\omega$
events.}
\label{g0ww}
\end{figure}

A MC simulation of $\psi(2S)\ar\gamma\chicJ,~\chicJ\ar\ww$ 
is used to determine the detection efficiency. The proper angular 
distributions of the photon emitted in $\psi(2S)\ar\gamma\chicJ$ are 
used~\cite{angle}.
Fig.~\ref{g0ww} shows the distributions, identical to those in Fig.~\ref{gww} 
for MC simulated $\psi(2S)\ar\gamma\chi_{c0},~\chi_{c0}\ar\omega\omega$ events 
passing the same selection criteria as for the real data. MC simulated 
$\psi(2S)\ar\gamma\chi_{c2},~\chi_{c2}\ar\omega\omega$ events 
have similar distributions.

\subsection{Mass spectrum fit}
The Maximum Likelihood (ML) method is used to fit the $M_{\ww}$ mass
spectrum of events in the   
$\omega$ pair 
signal region (Fig.~\ref{gww}(c)). The  $\chi_{0,2}$ signal functions are 
determined from MC simulation, as shown in Fig.~\ref{g0ww}(c) for $\chi_{c0}$,
while the background function is taken from the sideband distribution, 
shown in Fig~\ref{sb}(a). The fit result is represented by the solid curve in 
Fig.~\ref{gwwfit}, and the fit yields 
$$ N_{\chi_{c0}}=38.1\pm 9.6,~N_{\chi_{c2}}=27.7\pm 7.4. $$
\bfg[ht]
\vspace{-.5cm}
\centerline{\psfig{file=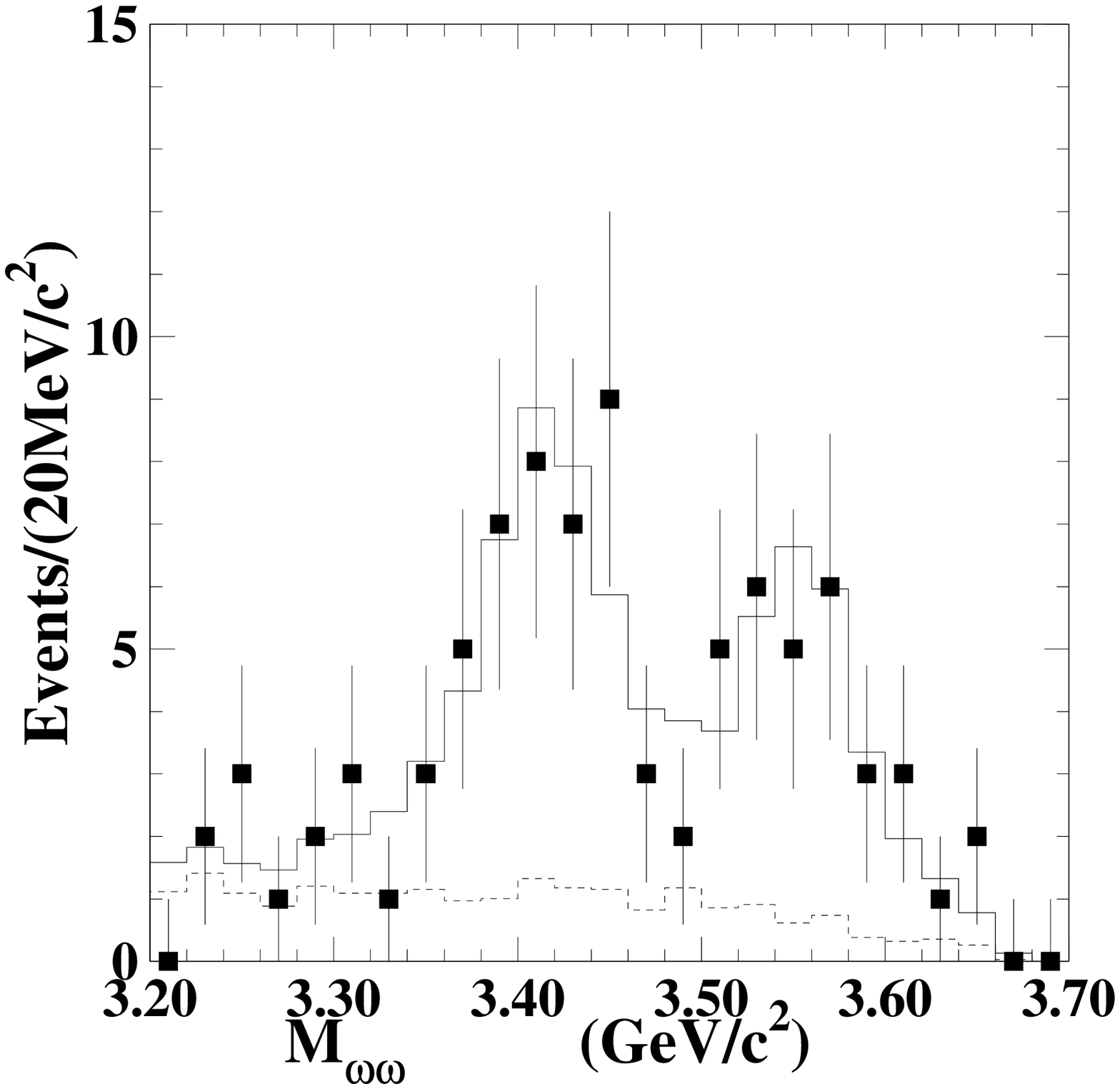,width=6.5cm,height=5.5cm}}
\caption{Fit of the $M_{\ww}$ distribution. Dots with error bars are
data, the solid histogram represents the maximum likelihood fit result, 
and the dashed histogram is the sideband background.}
\label{gwwfit}
\efg
The statistical significances of $\chi_{c0}$ and $\chi_{c2}$ are 
$4.4\sigma$ and $4.7\sigma$, respectively, which are estimated from 
$\sqrt{2\Delta ln{\cal L}}$, where $\Delta ln{\cal L}$ is the difference 
between the logarithmic ML values of the fit with and without the 
corresponding signal function.

\section{Systematic error}
The systematic error in this branching ratio measurement includes the
uncertainties in the MDC tracking efficiency, photon efficiency,
kinematic fit, background shape, number of $\psi(2S)$ events, etc.
\subsection{ MDC tracking efficiency and photon efficiency}
For charged tracks, the uncertainty of the tracking efficiency is
determined by comparing data and MC for some very clean $J/\psi$ decay
channels~\cite{simbes}, and an error of 2\% is found for
each track. A similar comparison has also been performed for
photons~\cite{lism}, and the difference is also about 2\% for a single
photon.

\subsection{ Kinematic fit}
The systematic error associated with the kinematic fit is due to
differences between data and MC simulation in the
determination of the track momentum, the track fitting error matrix, and the
photon energy and direction. The effect is studied for charged tracks
and neutral tracks separately. By comparing the number of events before
and after the kinematic fit for very clean event samples for data and
MC simulated data, the difference is determined to be 8.4\%, which is
taken as the systematic error.

\subsection{ Background shape}
Two different sideband $M_{6\pi}$ spectrum shapes, shown in Fig.~\ref{sb}, are 
used as the background function. The difference in the number of $\chi_{c0,2}$ events 
obtained with the two different shapes is taken as a systematic error.
\begin{table}[htpb]
\caption{Individual sources and total systematic error (\%).}
\bcl
\doublerulesep 2pt
\begin{tabular}{lcc}
\hline\hline
Source&$\chi_{c0}\ar\ww$&$\chi_{c2}\ar\ww$\\\hline\hline
track efficiency &8&8\\
photon efficiency &10&10\\
6-C fit &8.4&8.4\\
background shape &6.0&1.0\\
signal region &3.4&4.3\\
binning and fit range&1.4&3.2\\
angular distribution & -- &9.4\\
No. of $\psi(2S)$ events &4&4\\
${\cal B}(\psi(2S)\ar\gamma\chi_{cJ})$&5.1&6.7\\
${\cal B}(\omega\ar3\pi)$&0.9&0.9\\
${\cal B}(\pi^0\ar\GG)$& 0.0 & 0.0\\\hline
Total&18.1&20.4\\\hline\hline
\end{tabular}
\label{tot}
\ecl
\end{table}

\subsection{ Binning, fit range, and signal region}
The differences caused by different binning and fit ranges in the $\ww$ mass 
spectrum fit are 1.2\% and 3.4\% for $\chi_{c0}$ and $\chi_{c2}$, respectively. 
Different sized signal regions yield differences of 3.1\% and 2.4\% for 
$\chi_{c0}$ and $\chi_{c2}$, respectively, which are taken as a
systematic error.

\subsection{\boldmath Angular distribution of $\chi_{cJ}\ar\ww$}
In the estimation of the efficiency, a phase space generator
with only the angular distribution of the radiative photon
is considered. While this is correct for $\chi_{c0}$ decays, it
may introduce bias for $\chi_{c2}$ decays. The effect is estimated
by generating different angular distributions of the omega
in the $\chi_{c2}$ rest frame. The efficiency difference between
these tests and the phase space generator is estimated to be
9.4\%, which is put into the systematic error.
\subsection{Branching ratios of intermediate states}
The errors on intermediate state branching ratios are obtained from 
the PDG~\cite{pdg} except for ${\cal B}(\psi(2S)\ar\gamma\chicJ)$, where 
recent CLEO results~\cite{cleoc} are used.
Table~\ref{tot} summarizes all contributions to the systematic errors,
and the total systematic error is determined by the quadratic sum of all
terms.
\begin{table*}[htpb]
\caption{Branching ratio results and relevant numbers.}
\bcl
\doublerulesep 2pt
\begin{tabular}{lcc}\hline
Quantity&$\chi_{c0}\ar\ww$&$\chi_{c2}\ar\ww$\\\hline\hline
number of events &$38.1\pm 9.6$&$27.7\pm 7.4$\\
efficiency (\%)&1.66&1.55\\
$N_{\psi(2S)}~(\times 10^6)$&$14.00\pm 0.56$&$14.00\pm 0.56$\\\hline
${\cal B}(\omega\ar\ppp)~(\%)$&$89.1\pm 0.7$&$89.1\pm 0.7$\\
${\cal B}(\pi^0\ar\GG)~(\%)$&$98.798\pm 0.032$&$98.798\pm 0.032$\\
${\cal B}(\psi(2S)\ar\chicJ)~(\%)$&$9.22\pm 0.47$&$9.33\pm 0.63$\\\hline
${\cal B}(\chicJ\ar\ww)\cdot{\cal B}(\psi(2S)\ar\chicJ)~(\times
10^{-4})$&$2.12\pm 0.53\pm 0.37$&$1.65\pm 0.44\pm 0.32$\\
${\cal B}(\chicJ\ar\ww)~(\times 10^{-3})$&$2.29\pm 0.58\pm 0.41$&$1.77\pm
0.47\pm 0.36$\\\hline\hline
\end{tabular}
\label{result}
\ecl
\end{table*}
\section{Results}
The branching ratio of  ${\cal B}(\chicJ\ar\ww)$ is  determined from
$${\cal B}(\chicJ\ar\ww)=\frac{N^{obs}_{\chicJ}}{N_{\psi(2S)}\cdot
f_1\cdot f_2^2\cdot f_3^2\cdot\epsilon}$$
where $N^{obs}_{\chicJ}$ is the number of events selected, 
$N_{\psi(2S)}$ the total number of $\psi(2S)$ events,  $\epsilon$
is the detection efficiency for the investigated channel, and $f_1,f_2$
and $f_3$ are the branching ratios of
$\psi(2S)\ar\gamma\chicJ,\omega\ar 3\pi$, and $\pi^0\ar\GG$, respectively.
Table~\ref{result} 
lists the  $\chi_{c0,2}\ar\ww$ branching ratio results, together with  
numbers used in the branching ratio calculation.

In summary, $\ww$ signals in the decay of $\chi_{c0,2}$ are observed,
and their branching ratios measured for the first time. 
$\chi_{c0}$ and $\chi_{c2}$ decays to $\ww$ have similar decay branching
ratios, which is different from other
$\chicJ\ar VV$ decays $(\chicJ\ar\phi\phi,~\bar{K}^*(892)^0K^*(892)^0)$.
This measurement, together with previous measurements of $\chicJ\ar
VV$, will be helpful in understanding the nature of $\chicJ$ states.

\section{Acknowlegements}   
The BES collaboration thanks the staff of BEPC for their hard
efforts. This work is supported in part by the National Natural
Science Foundation of China under contracts Nos. 10491300,
10225524, 10225525, 10425523, the Chinese Academy of Sciences under
contract No. KJ 95T-03, the 100 Talents Program of CAS under
Contract Nos. U-11, U-24, U-25, and the Knowledge Innovation
Project of CAS under Contract Nos. U-602, U-34 (IHEP), the
National Natural Science Foundation of China under Contract No.
10225522 (Tsinghua University), and the Department of Energy under
Contract No.DE-FG02-04ER41291 (U Hawaii).

\end{document}